\documentclass[twocolumn,showpacs,aps,prl]{revtex4}
\usepackage{graphicx}
\usepackage{dcolumn}
\usepackage{amsmath}
\usepackage{latexsym}

\begin {document}

\title {Percolation model with an additional source of disorder}
\author
{Sumanta Kundu and S. S. Manna}
\affiliation
{
\begin {tabular}{c}
Satyendra Nath Bose National Centre for Basic Sciences,
Block-JD, Sector-III, Salt Lake, Kolkata-700106, India
\end{tabular}
}
\begin{abstract}
      The ranges of transmission of the mobiles in a Mobile Ad-hoc Network are not uniform in reality.
   They are affected by the temperature fluctuation in air, obstruction due to the solid objects, even the humidity 
   difference in the environment, etc. How the varying range of transmission of the individual active
   elements affects the global connectivity in the network may be an important practical question to ask.
   Here a new model of percolation phenomena, with an additional source of disorder, has been introduced for a
   theoretical understanding of this problem. 
   
      As in ordinary percolation, sites of a square lattice are occupied randomly with the probability $p$. 
   Each occupied site is then assigned a circular disc of random value $R$ for its radius. A bond is defined to be occupied 
   if and only if the radii $R_1$ and $R_2$ of the discs centered at the ends satisfy certain pre-defined condition. In a very
   general formulation, one divides the $R_1 - R_2$ plane into two regions by an arbitrary closed curve. One 
   defines that a point within one region represents an occupied bond, otherwise it is a vacant bond. Study
   of three different rules under this general formulation, indicates that the percolation threshold is always 
   larger and varies continuously. This threshold has two limiting values, one is $p_c$(sq), the percolation
   threshold for the ordinary site percolation on the square lattice and the other being unity. The variation
   of the thresholds are characterized by exponents, which are not known in the literature. In a special case, 
   all lattice sites are occupied by discs of random radii $R \in \{0,R_0\}$ and a percolation transition is 
   observed with $R_0$ as the control variable, similar to the site occupation probability. 
\end{abstract}

\maketitle

      A simple way to describe the phenomenon of percolation is to consider a rectangular slab 
   of porous material placed horizontally, and ask, if some liquid is poured on the top surface, will it 
   appear at the bottom surface? The answer is `yes' (`no'), depending on if the 
   fraction $p$ of the porous volume is larger (smaller) than a threshold value $p_c$ of the porosity \cite 
   {Stauffer,Grimmett,Sahimi}. It was Hammersley and Brodbent who introduced the percolation model by 
   occupying (pore space) randomly the sites of a regular lattice with probability $p$ and keeping them 
   vacant (rock matrix) with probability $(1-p)$ while trying to understand better the mechanism of
   gas masks \cite {Broadbent}. The percolation model can also be described by randomly occupying
   the bonds of the lattice. Till date, the percolation model is regarded as a simple model for studying the `order - 
   disorder' transition \cite {Sornette}. 

      Any two occupied sites (bonds), separated at a certain distance, are considered to be connected 
   if both belong to the same cluster of occupied sites (bonds). The correlation between them decreases
   with their distance of separation, and the functional form is exponential when the distance is large.
   The length scale that characterizes such a form is known as the correlation length $\xi(p)$, which 
   diverges as $p$ approaches a critical value $p_c$, known as the percolation threshold, that marks the 
   transition point between the ordered and 
   disordered phases. The best value of $p_c$(sq) for site percolation on the square lattice is 
   0.59274605079210(2) \cite {Jacobsen} and 1/2 for the bond percolation \cite {Ziff-Wiki}.
   In both cases, the nature of transition is continuous and they belong to the same universality class.

      Over the years a number of variants of the percolation model have been studied \cite {Araujo}. In the Continuum 
   Percolation \cite {Meester,Stanley}, one finds the minimal density of equal sized overlapping Lilies, floating at 
   random positions on the water surface of a pond, such that an ant will be able to cross the pond walking on the 
   Lilies \cite {Grimmett}. In a Mobile ad hoc network (MANET) each node represents a mobile 
   phone with a fixed range of transmission that is capable of receiving as well as transmitting signals 
   \cite {MANET}. Depending on the value of the range there exists a critical density of Lilies or phones 
   where the long range correlation appears \cite {Stauffer}.

      Recently, it has been suggested that a discontinuous transition may be possible in a model of percolation 
   and termed it as the ``Explosive Percolation'' \cite {EP,Ziff,Manna,Herrmann}. Subsequently, it has been shown that, 
   though a class of such models show very sharp changes in their order parameters for finite 
   size systems and therefore appear like discontinuous transition, they indeed exhibit continuous transition in 
   the asymptotic limit of very large system sizes \cite {Costa,Nagler,Riordan,Lee}.
   
       Here, we introduce a very general formulation of the percolation model. Sites of a square lattice of size 
   $L \times L$ are occupied randomly using circular discs of random radii values $R$. The transmission range of 
   a mobile phone in MANET may be compared to the radius $R$ of a disc. This range is affected by the temperature 
   fluctuation in air, obstruction due to the solid objects, humidity difference in the environment, etc. and 
   therefore, assuming random values for the radii of the discs is a better description than using the identical
   discs. In this prescription, a bond is defined to be occupied if and only if the radii $R_1$ and $R_2$ of discs 
   centered at the ends satisfy certain pre-defined rule, otherwise it is vacant. Most generally, the $R_1 - R_2$ 
   plane is divided into two different regions by an arbitrary closed curve. Any point within one region represents 
   an occupied bond, otherwise it is a vacant one. The percolation thresholds are larger and varies continuously 
   between $p_c$(sq) and unity. 
   
      The radii $R$ of the discs are drawn from a uniform rectangular distribution $P(R)$ of half width 
   $W$ and the centre at $R=1/2+S$, where $S$ denotes the shift parameter. For the simulation, a random 
   number $r \in \{0,1\}$ from a uniform distribution is assigned at each lattice site to calculate
   $R = 1/2 + S + (2r-1)W$.

\begin{figure}[t]
\begin {tabular}{cc}
\includegraphics[width=3.8cm]{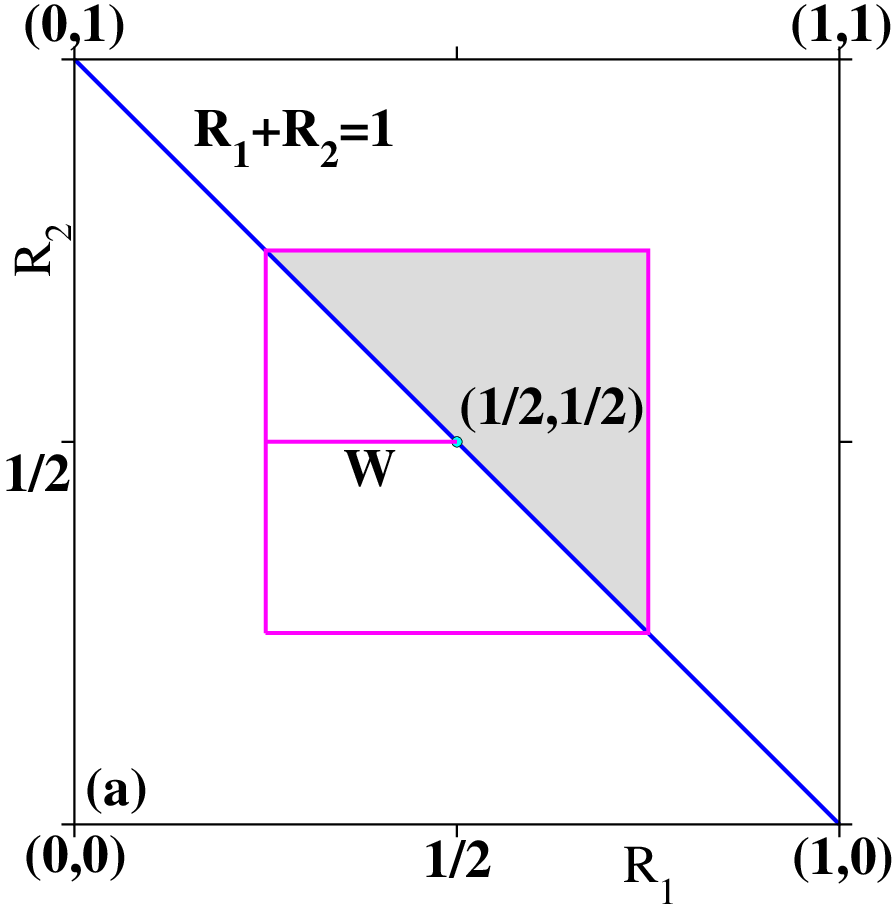} & \includegraphics[width=3.8cm]{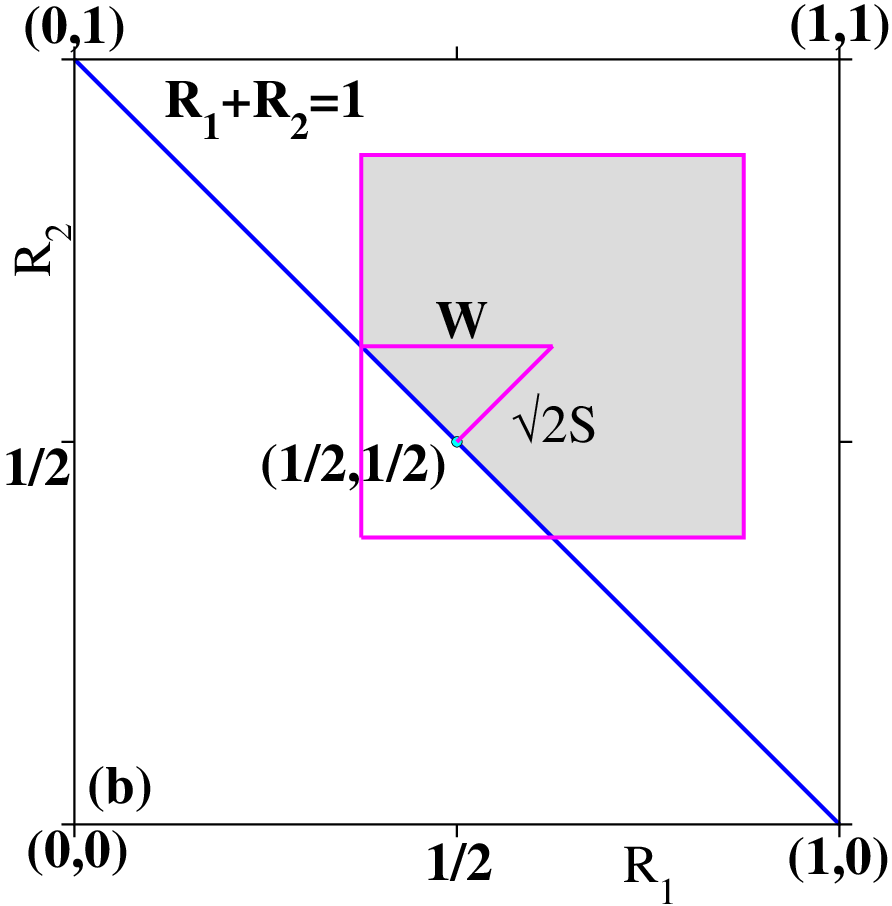} \\
\includegraphics[width=3.8cm]{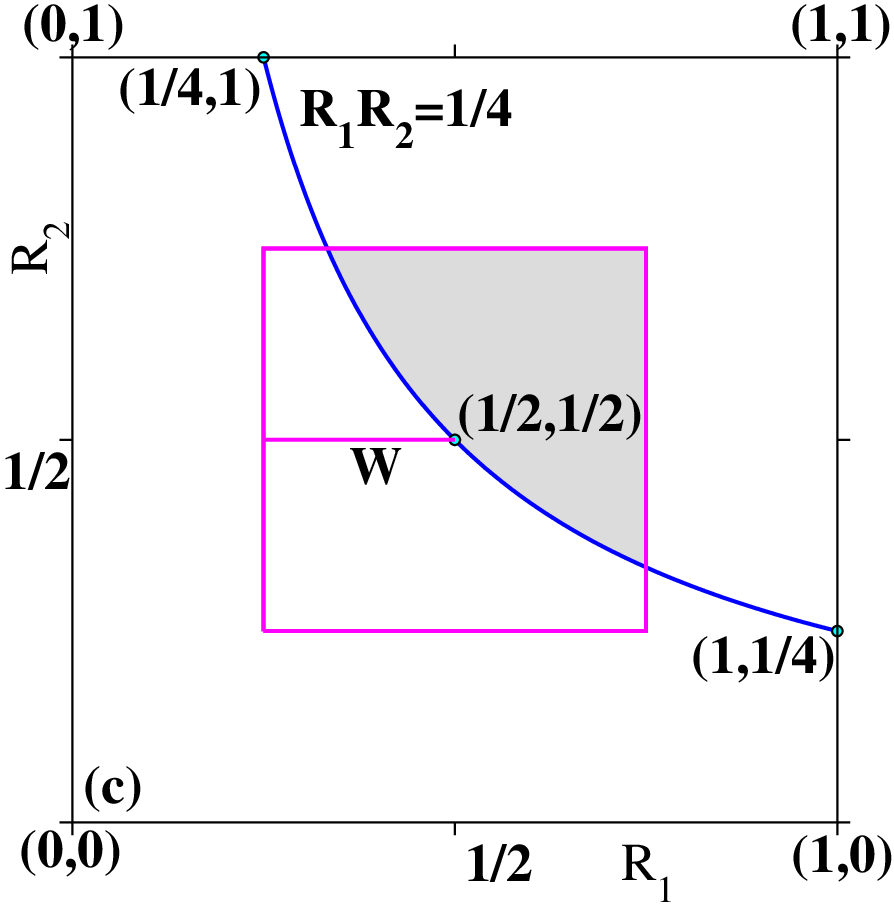} & \includegraphics[width=3.8cm]{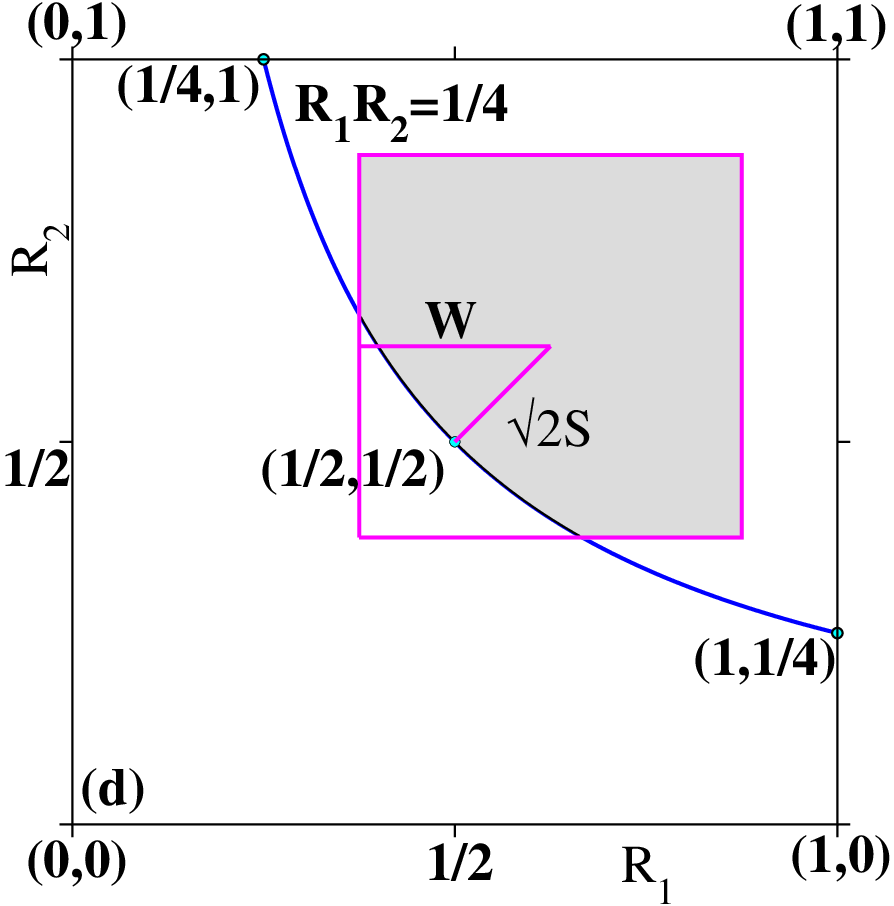} 
\end {tabular}
\caption{(Color online)
      On the $R_1-R_2$ plane, for a specific set of values of $W$=1/4 and $S$, the regions 
   corresponding to the occupied bonds (grey) and unoccupied bonds (white) are indicated. 
   Sum Rule: (a) $S$ = 0, (b) $S$ = 1/8 and the Product Rule: (c) $S$ = 0 and (d) $S$ = 1/8.
   }
\label {FIG01}
\end{figure}

\begin{figure}[t]
\begin {center}
\includegraphics[width=6.0cm]{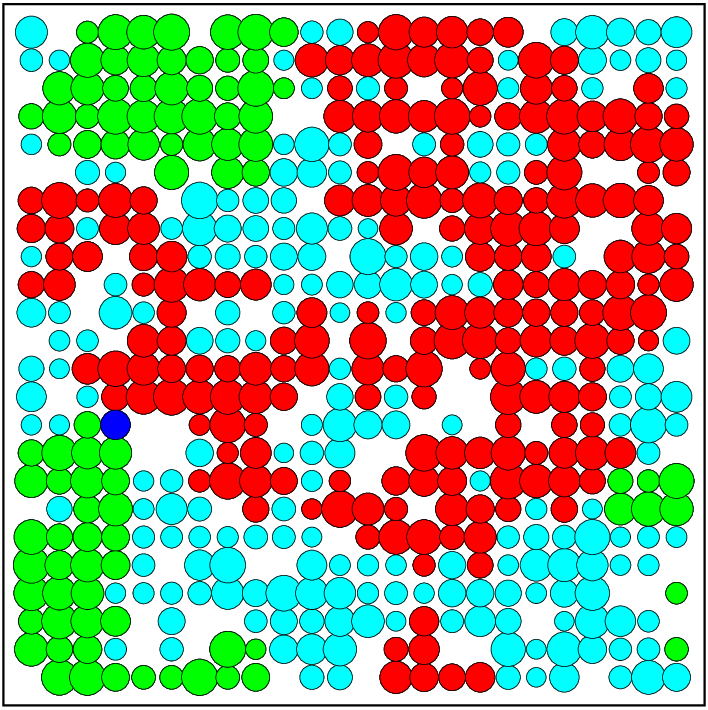}
\end {center}
\caption{(Color online)
A percolating configuration of 501 circular discs is drawn using the Sum Rule (with p.b.c.)
for $L$ = 24, $W$ = 0.15, $S$ = 0 and $p \approx$ 0.87. The largest and the second largest clusters are of sizes 
208 (red) and 90 (green) respectively. Because of the blue disc these two clusters merge and the maximal jump 
in the order parameter takes place. Discs at all other occupied sites are painted in cyan.}
\label {FIG02}
\end{figure}
   
   {\it The Sum Rule}: A bond is occupied, if and only if,
\begin {equation}
   R_1+R_2 \ge 1.
\label {EQN01}
\end {equation}

   For a given pair of $S$ and $W$, the points in the $R_1 - R_2$ plane, representing the 
   occupied and vacant bonds, lie within a square box (Fig. \ref {FIG01}). 
   In Fig. \ref {FIG01}(a) and (b) we exhibit two specific cases with $S$ = 0 and 1/8 respectively where $W$=1/4.
   A typical picture of a percolating configuration for the Sum Rule has been shown in Fig. \ref {FIG02}.

      To generate a single percolation configuration with the occupation probability $p$, we start
   from an empty square lattice of size $L$ and then drop $pL^2$ discs, one by one, on to the lattice 
   sites. At every step, an arbitrary site $i$ is randomly selected and if it is vacant, a disc with a randomly
   selected radius $R_i$ is placed at this site. Once $pL^2$ sites are occupied, all four neighboring bonds of 
   every occupied site are then tested for possible occupation. 
   The number of occupied bonds an occupied site may have, varies from 0 to 4 even if all neighboring sites are 
   occupied. In this way, all bonds are assigned their occupied / vacant status. A `cluster' is a set of occupied 
   sites interlinked by occupied bonds. A random configuration $\alpha$ has a number of clusters of different shapes and 
   sizes. The size $s$ of a cluster is the number of sites in the cluster and the size of the largest cluster is 
   denoted by $s^{\alpha}_{max}(p,L)$. The order parameter $\Omega(p,L)$ is defined by the configuration averaged fractional 
   size of the largest cluster, i.e., $\Omega(p,L) = \langle s^{\alpha}_{max}(p,L) \rangle/L^2$.

\begin{figure}[t]
\begin {center}
\includegraphics[width=6.0cm]{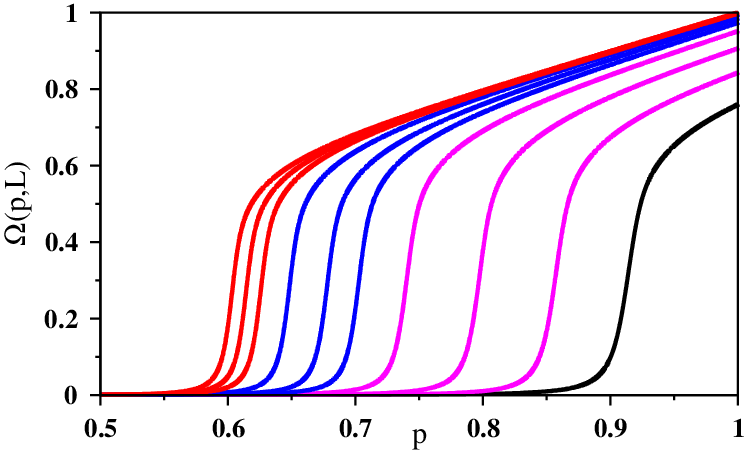}
\end {center}
\caption{(Color online) 
For the Sum Rule, the order parameter $\Omega(p,L)$ is plotted against the probability $p$ for $L$ = 512. 
 Colors used: red for $S$ = 0.03 and $W$ = 0.04, 0.045, 0.05; 
blue for $S$ = 0.02 and $W$ = 0.04, 0.05 , 0.06;
magenta for $S$ = 0.01 and $W$ = 0.04, 0.2/3, 0.15 and 
black for $S$ = 0 and $W$ = 0.04; the curves are arranged from left to right.
}
\label {FIG03}
\end{figure}
   
      By definition, as $p$ is gradually increased, the largest cluster grows monotonically. Around the 
   transition point, it makes several jumps in size when it merges with other clusters. For an arbitrary configuration,
   the largest cluster executes the maximal jump $\Delta_m s^{\alpha}_{max}(p,L)$ at $p=p^{\alpha}_c$, when it merges 
   with the maximal of the second largest cluster \cite {Margolina}. An average over many such configurations is considered as the 
   percolation threshold $p_c(L)=\langle p^{\alpha}_c \rangle$ for the system of size $L$. 
   
      For percolation model, it is well known that the correlation length diverges like $\xi(p) \propto |p_c-p|^{-\nu}$ 
   as $p \to p_c$ for the infinite system, where $\nu$ is the correlation length exponent and its value 
   is 4/3 in two dimension \cite {Stauffer,Eschbach}. However, for a finite size system $\xi$ may be at most $L$ and that is 
   attained at $p=p_c(L)$. Therefore, one gets $p_c(L) = p_c - AL^{-1/\nu}$ and the asymptotic value of $p_c$ is obtained by 
   extrapolating $p_c(L)$ against $L^{-1/\nu}$. It is also known that right at the percolation
   threshold the largest cluster is a fractal object, and its size grows as $\langle s^{\alpha}_{max}(p_c,L) \rangle \sim 
   L^{d_f}$, where $d_f$ is its fractal dimension in two dimension \cite {Feder}. Similarly, the maximal of the second largest cluster is
   also a fractal with the same fractal dimension $d_f$. As a consequence, the amount of the maximal jump in the order parameter 
   decreases with increasing $L$ as $\langle \Delta_m s^{\alpha}_{max}(p_c,L) \rangle/L^2 \sim L^{d_f-2}.$

      For $S$=0 and $W$=0, the bond between any pair of neighboring occupied sites is occupied. Therefore, $p_c(S=0,W=0)=p_c$(sq).
   When $W>0$, though only half of the discs have radii larger than 1/2, a global connectivity can still be achieved. The small 
   size discs certainly contribute to the density of occupied sites but may or may not take part in the bond density.    
   Consequently, it takes the higher density of occupied sites to attain the global connectivity. The growth of the largest cluster 
   is therefore retarded, i.e., $p_c(S=0,W>0) > p_c$(sq). Again because of the small discs, in the limit of $p \to 1$, the size 
   $s^{\alpha}_{max}(p,L)/L^2$ converges to a value which is well below unity, and it depends on the parameters $S$ and $W$.

\begin{figure}[t]
\begin {center}
\includegraphics[width=6.0cm]{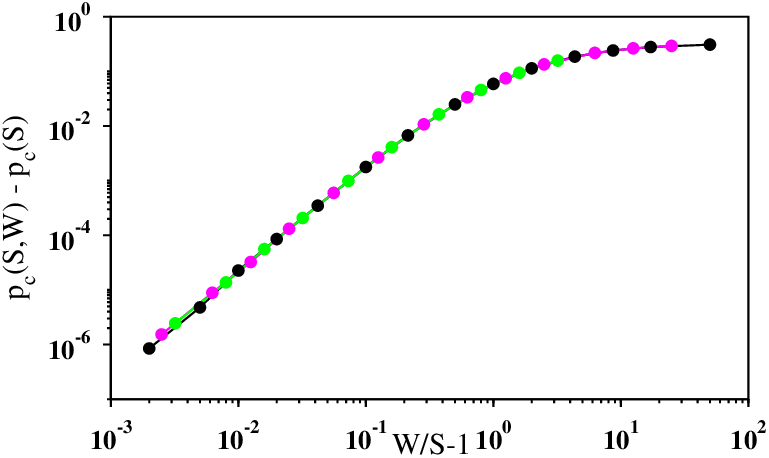}
\end {center}
\caption{(Color online)
    For the Sum Rule, the scaling plot of $p_c(S,W)-p_c(S)$ against $W/S-1$ has been shown
    for $S$ = 0.1 (green), 0.01 (magenta) and 0.001 (black). The values of $p_c(S)$ 
    required to make the curves straight in $W/S \to 1$ limit are 0.5927675, 0.5927684, and 
    0.5927662 respectively which are very close to $p_c$(sq). The slopes of the 
    linear portions are 1.96, 1.93 and 1.94 respectively, giving $\zeta_S=1.95(5)$.
}
\label {FIG04}
\end{figure}

      The $p_c(L)$ values are extrapolated against $L^{-1/\nu}$ with different trial values of $\nu$. The best fit corresponds to
   $\nu = 1/0.7502 \approx 1.333(5)$ and $p_c(S=0,W>0) \approx 0.9191(2)$. This is independent of $W$ since the bond occupation
   probability is 1/2 for all values of $W>0$. Secondly, the average fractional size of the largest cluster 
   has been found to decay like $L^{-0.105}$ and gives an estimate of $d_f=1.895(5)$ compared to the exact value of $d_f=91/48$ 
   \cite {Stauffer}. The average value of the maximal jump in the largest cluster varies as $L^{-0.104}$ and equating the power 
   to $d_f-2$ one gets $d_f$ = 1.896(5).
   
      Fig. \ref {FIG03} exhibits the variation of the order parameter $\Omega(p,L)$ against the site occupation probability $p$. 
   For $S$ = 0, the curve is independent of $W$. Further, for a fixed value of $S > 0$, the curve shifts to the higher values of $p$ 
   as $W$ increases, whereas, for a fixed value of $W$, the curve shifts towards the smaller values of $p$ as $S$ increases. 
   Numerically it appears that $p_c(L)$ depends only on ratio of $S$ and $W$. 
   
      For $S>0$, in the limit of $L \to \infty$, first the extrapolated values $p_c(S,W)$ are calculated. Then, a scaling analysis has 
   been done where we plot $p_c(S,W)-p_c(S)$ against $W/S -1$ in Fig. \ref {FIG04} and obtain a good data collapse. Tuning the values 
   of $p_c(S)$, the curves for different $S$ fit to a straight line as $W/S-1 \to 0$
   indicating a scaling form,
\begin {equation}
p_c(S,W) -p_c(S) \sim (W/S-1)^{\zeta_S}
\label {EQN08}
\end {equation}      
   where we estimated $\zeta_S = 1.95(5)$. The best tuned values of $p_c(S)$ are consistent with $p_c$(sq). 
   
      On the other hand, when $S$ is negative, the vacant area in Fig. \ref {FIG01}(a) increases, the occupied 
   area decreases and therefore the percolation threshold 
   increases. For a specific threshold value of $S=S_c= -0.0201(5)$ the $p_c(S_c)=1$ for $W=1/4$. It has been observed
   that $(p_c(S_c,W) - p_c(S,W)) \sim (S-S_c)^{\eta_S}$ with $\eta_S \approx 1.003(5)$. For other $W$ values $S_c(W)$ varies, 
   but $S_c(W)/W$ remains constant.

\begin{figure}[t]
\begin {center}
\includegraphics[width=6.0cm]{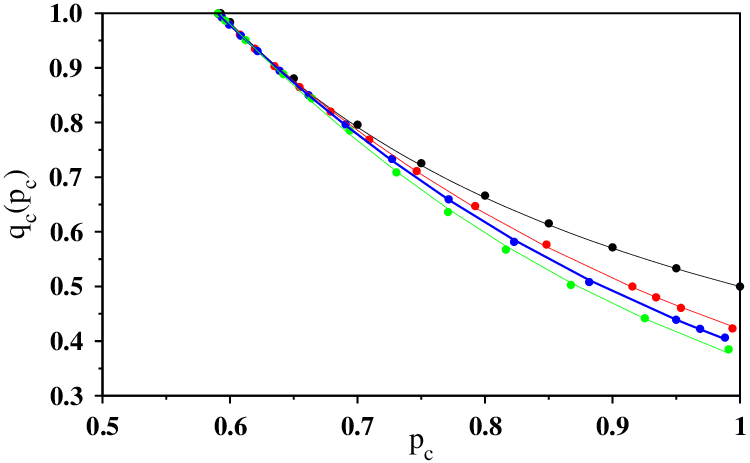}
\end {center}
\caption{(Color online)
   Critical values of the site $(p_c)$ and the bond $(q_c)$ occupation probabilities are plotted for 
   the site-bond percolation \cite {Tarasevitch} (black), Sum Rule (red), Product Rule (blue), and the Circular Rule (green).
   The solid lines are the best fitted forms given in Eqn. (4).
   }
\label {FIG05}
\end{figure}
   
\begin{figure}[t]
\begin {center}
\includegraphics[width=7.0cm]{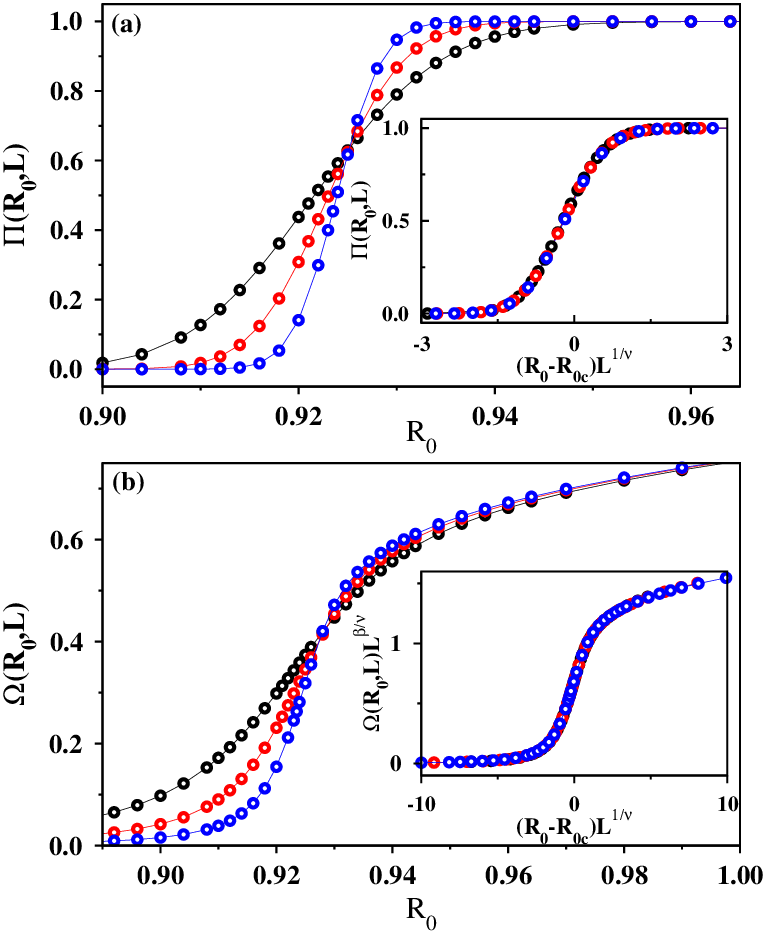}
\end {center}
\caption{(Color online) 
   For the system sizes $L$ = 256 (black), 512 (red), and 1024 (blue) and with $R_{0c}=0.925$, $1/\nu=0.75$ and $\beta/\nu=0.11$.
   (a) The percolation probability $\Pi(R_0,L)$ is plotted against $R_0$. 
       Inset: A scaling by $(R_0-R_{0c})L^{1/\nu}$ shows the data collapse.
   (b) The plot of order parameter $\Omega(R_0,L))$ against $R_0$. 
       Inset: A scaling by $\Omega(R_0,L))L^{\beta/\nu}$ against $(R_0-R_{0c})L^{1/\nu}$ exhibits an excellent data collapse.
}
\label {FIG06}
\end{figure}

     {\it The Product Rule}: Here, the condition for occupation of a bond is,
\begin {equation}
   R_1 R_2 \ge 1/4.
\label {EQN04}
\end {equation}
   Figs. \ref {FIG01}(c) and (d) represent occupied / vacant bonds determined by the Product Rule for $S$ = 0 and 
   1/8 respectively and with $W$=1/4.

      It can be seen from the Fig. \ref{FIG01}(c) that for $S$=0, the probability of an occupied bond (the shaded area) 
   for the Product Rule decreases with increasing $W$ and for this reason, the order parameter depends explicitly on 
   the value of the width $W$ and the critical percolation probability increases with $W$. On the other hand, for a 
   general value of $S>0$, the $\Omega(p,L)$ plots are quite similar to those of the Sum Rule, but $p_c(L)$ values are
   slightly larger. First, the asymptotic values of the critical percolation probabilities $p_c(S,W)$ for $S=0$ and 
   $W \to 0$ has again been found to be 0.9191(2). For $S > 0$, again a scaling plot of $p_c(S,W)-p_c(S)$ against $W/S-1$ 
   gives a very nice data collapse and we find $\zeta_P \approx 1.93(10)$. Here also the shift $S$ may take negative values 
   so that the percolation threshold would increase to unity i.e., $p_c(S_c,W)$=1 for $S_c = -0.0117(5)$ for $W$=1/4.
   The approach to this limit is again characterized by $\eta_P \approx 1$.

     {\it The Circular Rule}: Here a circular region, centered around the point $(1/2,1/2)$, of radius $\Delta$ in 
   the $R_1 - R_2$ plane is selected. The radii $R$ of the discs are again distributed by $P(R)$ but only 
   $S$ = 0 and $W$ = 1/2 are used. The region inside the circle represents the occupied bonds whereas the 
   outside region represents the vacant bonds.

     Evidently, the critical percolation threshold $p_c(\Delta,L)$ depends on the value of $\Delta$. 
   It has been observed that if the size of the circular region is too small, the size of the largest cluster
   becomes minuscule even when the occupation probability $p=1$. Consequently, one defines a threshold value 
   $\Delta_c$ such that a global percolation transition can occur only when $\Delta > \Delta_c$. Clearly, the critical 
   percolation probability at $\Delta_c$ is denoted by $p_c(\Delta_c)=1$. As before, $(p_c(\Delta_c)-p_c(\Delta))$ 
   varies as $(\Delta - \Delta_c)^{\eta_C}$. The best fitted value of $\Delta_c$ is found to be 0.3488(5) with 
   $\eta_C \approx 0.96(5)$. Also, the other limit corresponds to $\Delta_L=1/\sqrt 2$ when all points in the 
   $R_1 - R_2$ plane correspond to the occupied bonds. In this case ($p_c(\Delta) - p_c$(sq)) varies as 
   $(\Delta_L-\Delta)^{\zeta_C}$ and we estimated $\zeta_C \approx 1.95(5)$.

      Our model is distinctly different from the random site-bond percolation \cite {Coniglio,Tarasevitch}.
   In this model, sites and bonds of the same lattice are occupied independently. A connecting path is therefore a 
   sequence of alternate occupied sites and bonds and the global connectivity is determined by the appearance 
   of such paths across the system. In comparison, in our model when two neighboring sites are occupied, the 
   occupied / vacant status of the bond between them is immediately determined, subject to the fulfillment of
   certain condition.
   
      This difference shows up in the following example. In Fig. \ref {FIG01}(a), the grey area represents the 
   bond occupation probability $q$ = 1/2, where the percolation threshold is estimated to be $p_c \approx 0.9191$. 
   This is clearly different from the random site bond percolation on square lattice, which gives $p_c=1$ when 
   $q_c$ is set at 1/2 \cite {Tarasevitch}.
   
      In random site percolation, the bond density grows with the site density as $q(p)=p^2$. In comparison, in our case, 
   this form is modulated by a function as: $q(p)={\cal H}(S,W)p^2$ where, for the Sum Rule,
\begin {align}
& {\cal H}(S,W) = 1/2 + S/W - S^2/(2W^2), \text{for $S > 0$} \text{~~and} \nonumber \\
& {\cal H}(S,W) = 1/2 - S/W + S^2/(2W^2), \text{for $S < 0$}. \nonumber
\label {EQN09}
\end {align}
   For the Product Rule, there exists a threshold value $S_W$, such that for $S \le S_W$,
\begin {align}
& 4W^2{\cal H}(S,W) \nonumber \\
& =(S+W)^2+(S+W) - \ln(1+2S+2W)/2 \text{~~and} \nonumber \\
& 4W^2{\cal H}(S,W) \nonumber \\
& =4W^2-(S-W)^2-(S-W)+\ln(1+2S-2W)/2 \nonumber
\end {align}
for $S \ge S_W$ where, $S_W = [(1+4W^2)^{1/2} - 1]/2$. Our numerical estimations are very much consistent with
these expressions.

      In Fig. \ref {FIG05} we have shown the phase diagram, similar to the site-bond percolation. The phase space 
   in this diagram is divided into two regions, namely, the percolating and the non-percolating regions. Therefore, 
   every point on the boundary between the two regions signifies a critical point, represented by $(p_c,q_c(p_c))$.  
   The data for the random site-bond percolation have been collected from \cite {Tarasevitch}. Similar phase
   boundaries for the Sum, Product and the Circular rules have also been shown for comparison. All four phase
   boundaries are completely distinct in general, but they meet only at the point $(p_c$(sq),1). For the random 
   site-bond percolation, the functional form of the critical curve is $q_c(p_c) = B/(A+p_c)$ \cite {Tarasevitch}
   and is represented by the black solid line. Here we have tried a modified functional form to fit our data as:
\begin {equation}
q_c(p_c) = B/(A+p^{\theta}_c)
\end {equation}
   and we have observed that $\theta=$ 2.41, 2.70, and 2.81 for the Sum, Product and Circular rules respectively. 
   For the Sum and Product rules $W=1/4$ has been used.

      A very interesting special case of our model is the situation when all sites of he lattice are occupied ($p=1$)
   by discs of uniformly distributed radii $R \in \{0,R_0\}$. A related model in continuum percolation considers
   discs of randomly selected radii \cite {Lorenz,Quintanilla}. The set of occupied bonds are then determined by the 
   Sum Rule using the periodic boundary condition along the horizontal direction and the open boundary condition along 
   vertical direction. For any value of $R_0 < 1/2$, none of the bonds become occupied. When $R_0$ is further increased, 
   the size of the largest cluster exhibits a sharp increase, similar to the ordinary percolation, for a critical value 
   $R_{0c}$. We defined
   $\Pi(R_0,L)$ as the spanning probability from the top to the bottom of the lattice. We also calculated the order
   parameter $\Omega(R_0,L) = \langle s^{\alpha}_{max}(R_0,L) \rangle / L^2$.
   
      In Fig. \ref {FIG06}(a), we plot $\Pi(R_0,L)$ against $R_0$ for three different system sizes which 
   meet at approximately same value of $R_0 = R_{0c} = 0.925(5)$. A finite-size scaling of $\Pi(R_0,L)$ plotted against the 
   scaled variable $(R_0-R_{0c})L^{1/\nu}$ with $1/\nu=0.75$ works very well (Fig. \ref {FIG06}(a) inset), implying,
\begin {equation}
\Pi(R_0,L) \sim {\cal F}\big[(R_0-R_{0c})L^{1/\nu}\big].
\end {equation}
   Secondly, in Fig. \ref {FIG06}(b), we have plotted $\Omega(R_0,L)$ against $R_0$, and the scaling form
   (Fig. \ref {FIG06}(b) inset)
\begin {equation}
\Omega(R_0,L)L^{\beta/\nu} \sim {\cal G}\big[(R_0-R_{0c})L^{1/\nu}\big]
\end {equation}
   works excellent. Comparing with the ordinary percolation we recognize $\nu$ as the correlation length exponent and $\beta$ 
   as the order parameter exponent. Our best collapse of the data corresponds to $1/\nu=0.75$ and $\beta/\nu=0.110(5)$. These 
   values  are to be compared with the exact values of the two dimensional percolation exponents $\nu=4/3$ and $\beta=5/36$, 
   i.e., $\beta/\nu=5/48 \approx 0.1042$ \cite {Levinshtein, Margolina2}. The entire calculation has been repeated using the Product Rule and the results
   are found to be very similar to those of the Sum Rule except $R_{0c}$ = 0.978(5) and $\beta/\nu \approx 0.104(5)$.
 
      To summarize, in the Statistical Physics framework of the percolation phenomena we have attempted to study the 
   global connectivity problem in a Mobile Ad-hoc Network, where all active elements are not of uniform transmitting 
   capacities. Transmission ranges of different mobile elements may be different. Does the network still globally
   connected, is what we like to ask. Our theoretical study in this paper answers this question in the affirmative,
   which is also interesting from the point of view of critical phenomena of disordered systems.
      
      A very general percolation problem has been formulated with two different types of randomness. A bond is occupied 
   if the pair of 
   neighboring discs of randomly distributed radii $R_1$ and $R_2$ fulfills certain condition. Such a condition is
   most generally described by dividing the $R_1 - R_2$ plane into two regions by a closed curve of arbitrary shape; 
   one region represents the connected, where as the other region represents the vacant bonds. The percolation 
   threshold varies within $p_c$(sq) $ \le p_c \le 1$. The nature of the
   percolation transition is continuous, but the approach of the percolation threshold to its limiting values 
   is described in terms of new exponents $\zeta$ and $\eta$, not yet known in the literature. Moreover, our analysis 
   even on a fully occupied lattice reveals that a percolation transition can occur where the control parameter is 
   the maximal radius $R_0$ of the discs. The set of critical exponents exhibits excellent agreement with those of the ordinary 
   percolation, implying that both may belong to the same universality class.



\begin{thebibliography}{90}
\bibitem {Stauffer}    D. Stauffer and A. Aharony, {\it Introduction to Percolation Theory}, Taylor \& Francis, (2003).
\bibitem {Grimmett}    G. Grimmett, {\it Percolation}, Springer (1999).
\bibitem {Sahimi}      M. Sahimi. {\it Applications of Percolation Theory}, Taylor \& Francis, 1994. 
\bibitem {Broadbent}   S. Broadbent and J. Hammersley, {\it Percolation processes I. Crystals and mazes}, 
                       Proceedings of the Cambridge Philosophical Society {\bf 53}, 629 (1957).
\bibitem {Sornette}    D. Sornette, {\it Critical Phenomena in Natural Sciences: Chaos, Fractals, Selforganization 
         and Disorder: Concepts and Tools}, Springer (2006).
\bibitem {Jacobsen}    J. L. Jacobsen, J. Phys. A: Math. Theor., {\bf 48}, 454003 (2015).
\bibitem {Ziff-Wiki}   A complete list of percolation thresholds is in 
                       \begin{verbatim} en.wikipedia.org/wiki/Percolation_threshold. \end{verbatim}
\bibitem {Araujo}      N. Araujo, P. Grassberger, B. Kahng, K. J. Schrenk and R. M. Ziff, Eur. Phys. J. Special Topics {\bf 223},
                       2307 (2014).
\bibitem {Meester}     R. Meester and R. Roy, {\it Continuum Percolation}, Cambridge University Press, (1996).
\bibitem {Stanley}    E. T. Gawlinski and H. E. Stanley, J. Phys. A, {\bf 14}, L291 (1981).
\bibitem {MANET}       H. Mohammadi, E. N. Oskoee, M. Afsharchi, N. Yazdani, and M. Sahimi, Int. J. Mod. Phys. C 
                       {\bf 20}, 1871 (2009).
\bibitem {EP}          D. Achlioptas, R. M. D'Souza, and J. Spencer, Science {\bf 323}, 1453 (2009).
\bibitem {Ziff}        R. M. Ziff, Phys. Rev. Lett. {\bf 103}, 045701 (2009).
\bibitem {Manna}       S. S. Manna, Physica A, {\bf 391}, 2833 (2012).
\bibitem {Herrmann}    N. A. M. Araujo and H. J. Herrmann, Phys. Rev. Lett. {\bf 105}, 035701 (2010).
\bibitem {Costa}       R. A. da Costa, S. N. Dorogovtsev, A. V. Goltsev and J, F. F. Mendes, Phys. Rev. Lett., {\bf 105}, 255701 (2010).
\bibitem {Nagler}      J. Nagler, A. Levina, and M. Timme, Nature Physics, {\bf 7}, 265 (2011).
\bibitem {Riordan}     O. Riordan and L. Warnke, Ann. Appl. Prob. {\bf 22}, 1450 (2012).
\bibitem {Lee}         H. K. Lee, B. J. Kim, and H. Park, Phys. Rev. E {\bf 84}, 020101(R) (2011).
\bibitem {Margolina}   A. Margolina, H. J. Herrmann and D. Stauffer, Phys. Lett. {\bf 93A}, 73 (1982).
\bibitem {Eschbach}    P. D. Eschbach, D. Stauffer and H. J. Herrmann, Phys. Rev. B, {\bf 23}, 422 (1981).
\bibitem {Feder}       J. Feder, {\it Fractals}, Springer (1988).
\bibitem {Coniglio}    A. Coniglio, H.E. Stanley and W. Klein, Phys. Rev. Lett. {\bf 42}, 518 (1979).
\bibitem {Tarasevitch} Y. Y. Tarasevitch and S. C. Van der Marck, Int. J. Mod. Phys. C {\bf 10}, 1193 (1999).
\bibitem {Lorenz}      B. Lorenz, I. Orgzall, and H. O. Heuer, J. Phys. A, {\bf 26}, 4711 (1993).
\bibitem {Quintanilla} J. Quintanilla, Phys. Rev. E. {\bf 63}, 061108 (2001).
\bibitem {Levinshtein} M. E. Levinshtein, B. I. Shklovskii, M. S. Shur, and A. L. Efros, Zh. Eksp. Theor. Fiz., {\bf 69}, 386 (1975).
\bibitem {Margolina2}  A. Margolina and H. J. Herrmann, Phys. Lett. {\bf 104A}, 295 (1984).
\end {thebibliography}
\end {document}